\documentclass[12pt,aip,jap]{article}

\usepackage{cite}

\usepackage{graphicx}
\usepackage{subfigure}

\usepackage{amssymb}
\usepackage{amsmath}
\usepackage{amscd}
\usepackage{latexsym}

\newcommand{\be}{\begin{equation}}
\newcommand{\ee}{\end{equation}}
\newcommand{\Dlt}{\Delta}
\newcommand{\dlt}{\delta}

\newcommand{\br}{{\bf r}}

\newcommand{\bn}{{\bf n}}
\newcommand{\bbe}{{\bf e}}
\newcommand{\bB}{{\bf B}}
\newcommand{\bS}{{\bf S}}

\newcommand{\bt}{\beta}

\newcommand{\al}{\alpha}

\newcommand{\gm}{\gamma}
\newcommand{\om}{\omega}
\newcommand{\Om}{\Omega}
\newcommand{\Gm}{\Gamma}

\newcommand{\lgl}{\langle}
\newcommand{\rgl}{\rangle}

\begin{document}

\begin{center}

{\Large{\bf Fast magnetization reversal of nanoclusters in resonator } \\ [5mm]

V.I. Yukalov$^{1,}$\footnote{Electronic mail: yukalov@theor.jinr.ru} and E.P. Yukalova$^{2}$} \\ [3mm]

{\it 
$^1$Bogolubov Laboratory of Theoretical Physics, \\
Joint Institute for Nuclear Research, Dubna 141980, Russia \\ [3mm]

$^2$Laboratory of Information Technologies, \\
Joint Institute for Nuclear Research, Dubna 141980, Russia} 

\end{center}

\vskip 2cm

\begin{abstract}

An effective method for ultrafast magnetization reversal of nanoclusters is 
suggested. The method is based on coupling a nanocluster to a resonant electric 
circuit. This coupling causes the appearance of a magnetic feedback field 
acting on the cluster, which drastically shortens the magnetization reversal 
time. The influence of the resonator properties, nanocluster parameters, and 
external fields on the magnetization dynamics and reversal time is analyzed. 
The magnetization reversal time can be made many orders shorter than the 
natural relaxation time. The reversal is studied for both the cases of a 
single nanocluster as well as for the system of many nanoclusters interacting 
through dipole forces. 

\end{abstract}

\vskip 2cm


\newpage

\section{Introduction}

The effect of magnetization reversal in nanomaterials is of considerable 
importance for various magneto-electronic devices, magnetic recording and 
storage, and other information processing techniques. The standard way of 
recording an information bit is to reverse the magnetization by applying a 
magnetic field antiparallel to the magnetization. From another side, the 
nanoparticle magnetic moment has to be sufficiently stable, which can be 
achieved by the use of materials with high magnetic anisotropy. But the 
latter complicates the process of magnetization reversal. In order to 
resolve the contradiction between these two requirements, different methods 
of magnetization reversal have been suggested, as can be inferred from the 
review articles [1,2].

Magnetization reversals caused by thermal fluctuations and phonon-assisted 
quantum tunneling are rather slow processes at low temperatures, below the 
blocking temperature, where nanoclusters exhibit stable magnetization 
[1,2,3-10]. To make the reversal faster, several methods have been suggested. 
Thus, one can employ transverse magnetic constant fields [11] or short 
pulses [12-17], transverse microwave alternating fields at magnetic resonance 
frequency [18-24], and optical laser pulses [25].
 
In the present paper, we suggest another method for achieving an ultrafast
magnetization reversal of nanoclusters. The method is based on coupling the 
considered nanocluster with a resonator by placing the cluster into the 
magnetic coil of an electric circuit. Then the motion of the cluster 
magnetization produces a magnetic feedback field acting on the cluster. This 
feedback mechanism essentially accelerates the magnetization reversal. 
Actually, the effect of the accelerated thermalization of nuclear magnets 
was proposed by Purcell [26] and considered, using the classical Bloch 
equations, by Blombergen and Pound [27]. Here we study the magnetization 
reversal by employing quantum microscopic Hamiltonians typical of strongly 
anisotropic nanoclusters possessing large spins.        
 
We study magnetic clusters with the effective sizes shorter than the exchange 
length of atoms composing them, when the magnetic cluster is in a single-domain 
state and its magnetization can be represented by a large total spin. Such 
clusters are necessarily of nanosizes, which explains their name of nanoclusters.  
To avoid complications, due to distributions of particle sizes and shapes, we 
consider the magnetization dynamics of similar nanoclusters. There are two 
essentially different cases. One is the magnetization reversal of a single 
nanocluster. And the other is the magnetization dynamics in an ensemble of 
nanoclusters interacting through dipolar forces. We study both these limiting 
cases.

\section{Spin Dynamics of Magnetic Nanoclusters}

Let us, first, consider a single nanocluster. The typical Hamiltonian of a 
nanocluster, with the total spin ${\bf S}$, can be written in the form 
\be
\label{1}
 \hat H = -\mu_0 \bB \cdot \bS - D \left ( S^z \right )^2 +
D_2 \left ( S^x \right )^2 \;  ,
\ee
where $\mu_0 = - \hbar \gamma_S$ is the cluster magnetic moment, 
$\gamma_S \approx 2\mu_B/\hbar$ is the gyromagnetic ratio, $\mu_B$, Bohr magneton, 
${\bf B}$ is the total magnetic field acting on the cluster, $D$ and $D_2$
are anisotropy constants. This Hamiltonian reminds the classical 
Stoner-Wohlfarth model [28,29]. However, we start with the microscopic quantum 
Hamiltonian (1), where the spin vector is treated as an operator. This will
allow us to explicitly define all system parameters and to take into account 
quantum effects that can be important for the dynamics of nanocluster 
assemblies.

One usually represents the cluster energy in a reduced form [1,2] with the 
anisotropy parameters related to $D$ and $D_2$ as
\be
\label{2}
 K = \frac{DS^2}{V_1} \;  , \qquad K_2 = \frac{D_2 S^2}{V_1} \; ,
\ee
with $V_1$ being the single-cluster volume and $S$, the cluster spin value. 
The second-order magnetic anisotropy is caused by magnetocrystalline anisotropy,
shape anisotropy, and surface anisotropy [1,2]. Sometimes, one includes the
fourth-order and sixth-order anisotropy. However such higher-order anisotropy 
terms are usually much smaller than the second-order terms, so their inclusion
does not essentially change the overall picture.

The total magnetic field is the sum
\be
\label{3}
\bB = B_0 \bbe_z + H \bbe_x + B_1 \bbe_y
\ee
of an external constant magnetic field, generally having the longitudinal, 
$B_0$, and transverse, $B_1$, components, and of the resonator feedback field
$H$ directed along the axis of the coil, where the cluster is inserted to.
The field $B_0$ is directed opposite to the initial cluster magnetization. 
The magnetic field, created by the coil, is described by the Kirchhoff equation. 
The latter, as is known, defines electric current generated in the coil by 
varying magnetization. In turn, the current produces magnetic field along the 
coil axis. The resulting equation for the generated magnetic field $H = H(t)$ 
can be written [30,31] in the form  
\be
\label{4}
 \frac{dH}{dt} + 2 \gm H + \om^2 \int_0^t H(t') \; dt' =
- 4\pi\eta \; \frac{dm_x}{dt} \;  ,
\ee
in which $\omega$ is the circuit natural frequency, $\gamma$ is the circuit 
attenuation, $\eta = V_1/V_{res}$ is the filling factor, with $V_{res}$
being the volume of the resonant coil, and 
\be
\label{5}
m_x \equiv \frac{\mu_0}{V_1 } \; \lgl S^x \rgl
\ee
is the average magnetization density, corresponding to the mean spin in the 
direction of the coil axis. Thus, moving spins produce magnetic field that acts 
back on spins, accelerating their motion.
  
To write down the equations of motion for the spin operators, it is convenient 
to introduce several notations. We define the Zeeman frequency
\be
\label{6}
 \om_0 \equiv - \; \frac{\mu_0}{\hbar} \; B_0 = \gm_S B_0 \;  ,
\ee
the transverse frequency
\be
\label{7}
\om_1 \equiv - \frac{\mu_0}{\hbar}\; B_1 = \gm_S B_1 \;   ,
\ee
and the effective transverse field acting on the spin,
\be
\label{8}
 f \equiv -\; \frac{i}{\hbar} \; \mu_0 H + \om_1 \; .
\ee
Then the Heisenberg equations of motion for the spin operators 
$S^{\pm}=S^x\pm S^y$ take the form
\be
\label{9}     
 \frac{dS^-}{dt} = - i \om_0 S^- + f S^z + i\; \frac{D}{\hbar}\;
\left ( S^- S^z + S^z S^- \right ) + i\; \frac{D_2}{\hbar}\;
\left ( S^x S^z + S^z S^x \right ) \; ,
\ee
plus its Hermitian conjugate. And the equation for the longitudinal spin becomes
\be
\label{10}
 \frac{dS^z}{dt} = - \; \frac{1}{2} \; \left ( f^* S^- + S^+ f \right ) +
 \frac{D_2}{\hbar}\;  \left ( S^x S^y + S^y S^x \right ) \; .
\ee

The measurable quantities are the average spin components, such as the reduced 
transverse component
\be
\label{11}
u \equiv \frac{1}{S} \; \lgl S^- \rgl
\ee
and the reduced longitudinal magnetization
\be
\label{12}
 s \equiv  \frac{1}{S} \; \lgl S^z \rgl \; .
\ee

Averaging the equations of motion (9) and (10), we use the decoupling
\be
\label{13} 
 \lgl S^\al S^\bt + S^\bt S^\al \rgl = \left ( 2 \; - \; \frac{1}{S}
\right ) \lgl S^\al \rgl \lgl S^\bt \rgl \qquad (\al \neq \bt) \; ,
\ee
which preserves the correct behavior for all spins. Thus, for spin 
one-half it gives exactly zero, because of the anticommutativity of different 
Pauli matrices, and it becomes asymptotically exact for large spins [32-34]. 
We also take into account that nanoclusters possess a longitudinal 
relaxation rate $\gamma_1$ that is due to phonon-assisted tunneling and to 
spin-phonon interactions of the nanocluster with its surrounding. 

Let us introduce the longitudinal anisotropy frequency
\be
\label{14}
  \om_D \equiv ( 2S - 1)\; \frac{D}{\hbar} \; ,
\ee
the transverse anisotropy frequency
\be
\label{15}
 \om_2 \equiv ( 2S - 1)\; \frac{D_2}{\hbar} \;  ,
\ee
the effective anisotropy frequency
\be
\label{16}
  \om_A \equiv \om_D + \frac{1}{2} \; \om_2 \; ,
\ee
and the effective rotation frequency 
\be
\label{17}
  \Om \equiv \om_0 - \om_A s \; .
\ee
Also, let us define the effective field
\be
\label{18}
 F \equiv f + \frac{i}{2} \; \om_2 u^* = - \; \frac{i}{\hbar} \;
\mu_0 H + \om_1 + \frac{i}{2} \; \om_2 u^* \;  .
\ee
Then averaging Eqs. (9) and (10) yields the equations for the transverse 
spin component,
\be
\label{19}
\frac{du}{dt} = - i \Om u + Fs \;  ,
\ee
plus its complex conjugate, and for the spin magnetization,
\be
\label{20}
\frac{ds}{dt} = - \; \frac{1}{2} \left ( u^* F + F^* u \right ) -
\gm_1 ( s - \zeta ) \;   ,
\ee
where $\zeta$ is the equilibrium magnetization of a cluster.

There exists a large variety of different nanoclusters [1,2,6,35-38], 
because of which the system parameters can take values in a wide range. 
Concrete examples will be discussed in the concluding section.

\section{Approximate Analysis of Spin Dynamics}

The equations of motion (19) and (20) are derived for arbitrary cluster
parameters. Their solution can be done numerically. But before we pass
to numerical investigation, it is useful to give an approximate qualitative 
analysis allowing for the better understanding of physics involved. For this
purpose, we shall assume that some of the parameters are small as compared to 
the Zeeman and resonator frequencies.

First, it is possible to show [32-34] that the feedback equation (4) can be
represented as the integral equation
\be
\label{21}
 H = - 4\pi \eta \int_0^t G(t-t') \dot{m}_x(t') \; dt' \;  ,
\ee
in which the transfer function is
$$
 G(t) = \left [ \cos(\widetilde\om t) \; - \; \frac{\gm}{\widetilde\om}\;
\sin(\widetilde\om t) \right ] \; e^{-\gm t} \;  ,
$$
with the shifted frequency 
$$
 \widetilde\om \; \equiv \; \sqrt{\om^2 -\gm^2} \;,
$$
and the source is
$$
\dot{m}_x = \frac{\mu_0 S}{2V_1 } \; \frac{d}{dt}\;
\left ( u^* + u \right ) \;   .
$$

The resonator feedback field efficiently acts on the cluster only when the 
resonator natural frequency is tuned close to the Zeeman frequency, so that
the detuning be small, satisfying the resonance condition
\be
\label{22}  
  \left | \frac{\Dlt}{\om} \right | \ll 1 \qquad 
( \Dlt \equiv \om-\om_0 ) \; .
\ee
The effective rotation frequency (17) should also be close to $\omega$, which 
requires that the anisotropy frequency (16) be also sufficiently small,
\be
\label{23}
 \left | \frac{\om_A}{\om} \right | \ll 1  .
\ee
We assume that all attenuation parameters are smaller than $\omega$, such that
\be
\label{24}
 \frac{\gm}{\om}  \ll 1  \; , \qquad \frac{\gm_1}{\om}  \ll 1  \; .
\ee
And let us introduce the parameter playing the role of the {\it feedback rate}
\be
\label{25}
 \gm_0 \equiv \pi \eta \; \frac{\mu_0^2 S}{\hbar V_1} = 
\pi \; \frac{\mu_0^2 S}{\hbar V_{res}}  \; , 
\ee
which characterizes the attenuation caused by the coupling between the cluster
and resonator. 

Under these conditions, the solution to the integral equation (21) can be found
by an iterative procedure [32-34], which here gives
\be
\label{26}
 \mu_0 H = i \hbar \left ( \al u - \al^* u^* \right ) \; ,
\ee
where the {\it coupling function}, for small detuning, such that $|\Dlt|<\gm$,
reads as  
\be
\label{27}
 \al = \frac{\gm\gm_0 \Om}{\gm^2+\Dlt^2} \; \left ( 1 \; - \;
e^{-\gm t} \right ) \;  .
\ee

Substituting form (26) into Eqs. (19) and (20) results in the equations
$$
\frac{du}{dt} = - i \Om u + \al s u + \om_1 s -  
s \left ( \widetilde\al u \right )^* \; ,
$$
$$
\frac{dw}{dt} = 2\al s w + \om_1 \left (u^* + u \right ) s -
2s {\rm Re} \left ( \widetilde\al u^2 \right ) \; ,
$$
\be
\label{28}
\frac{ds}{dt} = -\al w \; - \; \frac{\om_1}{2} \; \left (u^* + u \right )
-\gm_1 ( s - \zeta ) +  {\rm Re} \left ( \widetilde\al u^2 \right ) \;  ,
\ee
in which
\be
\label{29}
w \equiv |\; u \; |^2 \; , \qquad  
\widetilde\al \equiv \al + \frac{i}{2} \; \om_2 \;  .
\ee

According to inequalities (22) to (24), the variables $w$ and $s$ can be treated
as slowly varying in time, while $u$ as a fast variable. The existence of two 
time scales allows us to invoke the scale separation approach [30,31,39,40] that 
is a variant of the averaging techniques [41]. Then we solve the equation for $u$, 
keeping there $w$ and $s$ as quasi-integrals of motion, which gives
\be
\label{30}
u = - \; \frac{i\om_1 s}{\Om+i\al s} + \left ( u_0 +
\frac{\om_1 s}{\Om+ i\al s} \right ) 
\exp \left \{ - i ( \Om + i \al s ) t \right \} \;   ,
\ee
where $u_0 \equiv u(0)$. This solution is to be substituted into the equations 
for the slow variables $w$ and $s$, with averaging their right-hand sides over 
time, thus, obtaining the equations for the guiding centers of $w$ and $s$. 

The scale separation approach is a very powerful method for dealing with a 
system of many interacting elements, such as the assemblies of spins [30-34] or
of quantum dots [42]. However, it requires the use of limitations on the range
of the system parameters, such as inequalities (22) to (24). While, in our case, 
we have derived Eqs. (19) and (20) that are valid for arbitrary parameter values. 
The latter equations are not difficult to solve numerically. But the present 
approximate consideration is useful for understanding the following physical 
points.   

The source of creating the resonator feedback field (21) is the motion of the 
cluster spin, which generates the magnetic field (26) acting back on the spin.
The back action is absent at the initial time. The coupling of the cluster spin 
and the resonator increases with time according to the behavior of the coupling 
function (27). The transverse spin component (30) oscillates with time with
the rotation frequency (17). The initial push to the spin oscillation is done
by the transverse field $B_1$ entering through the transverse frequency (7).
This also requires that the initial magnetization (12) be nonzero. Generally, 
there is one more source triggering the initial spin motion. These are quantum 
spin fluctuations that start the spin motion even in the absence of a transverse 
field, as has been shown for multi-spin systems [30-33]. Such quantum spin 
fluctuations are especially important for particles with low spins. The input 
of quantum fluctuations for large spins $S \gg 1$ diminishes as $1/S$, in 
agreement with Eq. (13). Therefore, in the case of a single nanocluster, these 
fluctuations can be neglected as soon as the transverse field $B_1$ is present. 
The transverse field, that can be easily regulated, serves as a convenient 
triggering mechanism for initiating spin motion.

\section{Numerical Solution of Evolution Equations}

Now we go back to the general evolution equations (19) and (20). For numerical
investigation, it is appropriate to pass to dimensionless quantities. To this 
end, we define the dimensionless feedback field
\be
\label{31}
 h \equiv - \; \frac{\mu_0 H}{\hbar\gm_0} = \frac{\gm_S H}{\gm_0} \;  .
\ee
The effective forces (8) and (18), respectively, become
\be
\label{32}
 f = i\gm_0 h + \om_1 \; , \qquad
F = i\gm_0 h + \frac{i}{2} \; \om_2 u^* + \om_1 \;  .
\ee

Instead of the complex variable (11), let us use the real components
\be
\label{33}
x \equiv \frac{1}{S} \; \lgl S^x \rgl \; , \qquad 
y \equiv \frac{1}{S} \; \lgl S^y \rgl \;   ,
\ee
when
$$
u = \frac{1}{S} \; \lgl S^- \rgl = x - i y \; .
$$
And in the following, let us measure time in units of $1/\gamma_0$.

With these notations, Eqs. (19) and (20) yield the equations for the spin 
components:
\be
\label{34}
 \frac{dx}{dt} = - \om_0 y + \om_D y s + \om_1 s \;  ,
\ee
\be
\label{35}
 \frac{dy}{dt} =  \om_0 x - ( \om_D + \om_2 ) x s - h s \;  ,
\ee
\be
\label{36}
 \frac{ds}{dt} = h y - \om_1 x + \om_2 x y  - \gm_1 ( s - \zeta ) \; .
\ee 
The feedback equation (4) can be rewritten in the form
\be
\label{37}
 \frac{d^2 h}{dt^2} + 2 \gm \; \frac{dh}{dt} + \om^2 h =
4 \; \frac{d^2 x}{dt^2} \;  .
\ee
We assume that at the starting moment of time the cluster magnetization 
is polarized along the axis $z$, which implies the initial spin components
\be
\label{38}
 x_0 \equiv x(0) = 0 \; , \qquad   y_0 \equiv y(0) = 0 \; , \qquad
 s_0 \equiv s(0) = 1 \; .
\ee
And for the initial feedback field we set
\be
\label{39}
  h_0 \equiv h(0) = 0 \; , \qquad  \dot{h}_0 \equiv \dot{h}(0) = 0 \; , 
\ee
where the overdot signifies time derivative. 

The overall time evolution is described by Eqs. (34) to (37), with the 
initial conditions (38) and (39). These equations possess the stable 
stationary solution 
\be
\label{40}
x^* = 0 \; , \qquad y^* =\frac{\om_1\zeta}{\om_0-\om_D\zeta} \; , \qquad
s^* = \zeta \; , \qquad h^* = \dot{h}^* = 0 \; ,
\ee
that is reached in the relaxation time $T_1 = 1/\gamma_1$. The attenuation
$\gamma_1$ can be defined by the Arrhenius law
$$
\gm_1 = \gm_A \exp \left ( - \; \frac{E_A}{k_B T} \right ) \;   ,
$$
where $E_A = \hbar \omega_A S$ is the anisotropy barrier. At low temperatures,
below the blocking temperature, the attenuation $\gamma_1$ is exponentially 
suppressed, so that the relaxation time $T_1$ is very long. However the 
magnetization reversal can be ultrafast due to the action of the resonator 
feedback field. In our calculations, we set $\gamma_1 = 10^{-3}$ 
(in units of $\gamma_0$) and take the initial conditions (38) and (39).

Figure 1 shows the solutions to the evolution equations (34) to (37), with 
the initial conditions (38) and (39), as functions of time 
(in units of $1/\gamma_0$) for typical parameters expressed in units of 
$\gamma_0$. The resonator feedback field realizes the magnetization reversal
that can be many orders shorter than the relaxation time $T_1$. Comparing
the behavior in the presence of the resonator and in its absence, we see
that in the time scale of $1/\gamma_0$ the influence of $\gamma_1$ is not 
noticeable at all.  

Figure 2 demonstrates the influence of the resonator attenuation on the 
magnetization reversal. Too short $\gamma$ means a large ringing time 
$1/\gamma$, when the resonator several times exchanges the energy with
the cluster, which leads to the oscillations of the magnetization. The 
optimal value of the resonator attenuation is $\gamma = 1$, when the 
resonator attenuation $\gamma$ coincides with the attenuation $\gamma_0$ 
characterizing the coupling between the resonator and the cluster.

Figure 3 illustrates the role of the Zeeman frequency. The larger the 
latter, the shorter the reversal time. However too large $\omega_0$
leads to many oscillations of magnetization, which is not good, if
one aims at the stable reversal.    

Figure 4 shows the influence of the transverse field $B_1$ entering through
the transverse frequency $\omega_1$. The larger the latter, the shorter the 
reversal time. But too large $\omega_1$ results in multiple oscillations,
which may be inconvenient for practical purposes.  

Figure 5 demonstrates the role of magnetic anisotropy. The anisotropy 
frequencies that are smaller than the Zeeman frequency do not strongly 
influence the reversal. But if the anisotropy frequencies are larger than
the Zeeman frequency, then the reversal is blocked. 

Figure 6 shows the importance of the resonance condition. Large detuning 
from resonance makes the magnetization reversal slower. The larger the 
detuning, the slower the reversal. The reversal is the fastest when the 
resonance condition $\omega_0 = \omega$ is valid.

These figures demonstrate that the coupling to a resonant electric circuit 
results in the ultrafast magnetization reversal of a single nanocluster, 
as compared to its natural relaxation time caused by thermal fluctuations and 
phonon-assisted tunneling. Estimates for typical nanoclusters will be given 
in the concluding section.

\section{System of Nanoclusters with Dipolar Interactions}

In the previous sections, we have treated the magnetization reversal of 
a single nanocluster. An important question is whether such an ultrafast 
magnetization reversal could be achieved for an ensemble of nanoclusters. 
The basic difference of the latter case from that of a single cluster is the 
existence of strong dipolar interactions between the nanoclusters. These 
dipolar interactions completely suppress coherent spin motion in dephasing 
time $T_2$, so that collective spin rotation, without a resonant feedback, 
becomes impossible [32-34,43]. One should not confuse the real dipolar 
interactions between spins with the effective interactions through photon 
exchange of resonant atoms radiating at optical frequencies. The dipolar 
spin interactions dephase spin motion, while the atomic interactions through 
the photon exchange, vice versa, collectivize atomic radiation [44]. Spin
motion, of course, also produces electromagnetic radiation that, however,
is extremely weak and can never collectivize spins in time shorter than the 
dephasing time [33,34,43,44]. Self-organized coherent atomic radiation,
called superradiance, is the Dicke effect [45]. The principally different
collectivization of spin motion by means of a resonator feedback field is
what is termed the Purcell effect [26]. The Dicke effect for spin systems
is impossible, so that the self-organized coherent spin motion is admissible 
only through the Purcell effect [32-34,43,44], which necessarily requires 
the coupling of the spin system with a resonator. 

Let us consider a system of $N$ nanoclusters in volume $V$, with the density
$\rho \equiv N/V$. The system Hamiltonian reads as
\be
\label{41}
\hat H = \sum_{j=1}^N \hat H_j + 
\frac{1}{2} \sum_{i\neq j}^N \hat H_{ij} \; .
\ee
Here the first term is the sum of the single-cluster Hamiltonians
\be
\label{42}
 \hat H_j = -\mu_0 \bB \cdot \bS_j - D(S_j^z)^2 + D_2 (S_j^x)^2 \;  ,
\ee
with the cluster spin operators ${\bf S}_j$ and the index $j = 1,2,\ldots,N$ 
enumerating the nanoclusters. Aiming at studying the role of the dipolar 
interactions, we keep in mind similar nanoclusters with close anisotropy 
parameters and spins $S$. The dipolar interactions are characterized by the 
Hamiltonian parts
\be
\label{43}
 \hat H_{ij} = \sum_{\al\bt} D_{ij}^{\al\bt} S_i^\al S_j^\bt \; ,
\ee
with the dipolar tensor
\be
\label{44}
 D_{ij}^{\al\bt} = \frac{\mu_0^2}{r_{ij}^3} \; \left ( \dlt_{\al\bt}
- 3 n_{ij}^\al n_{ij}^\bt \right ) \;  ,
\ee
in which
$$
 r_{ij} \equiv | \br_{ij} | \; , \qquad 
\bn_{ij} \equiv \frac{\br_{ij}}{r_{ij}} \; , \qquad 
\br_{ij} =\br_i - \br_j \; .
$$
The resonator feedback field is described by the same Eq. (4), but with
\be
\label{45}
 m_x = \frac{\mu_0}{V} \; \sum_{j=1}^N \; \lgl S_j^x \rgl \;   .
\ee

To write the equations of motion in a compact form, we introduce several 
notations. The dipolar terms are combined into the variables
\be
\label{46}
 \xi_i^0 \equiv \frac{1}{\hbar} \sum_{j(\neq i)}^N \left ( a_{ij} S_j^z +
c_{ij}^* S_j^- + c_{ij} S_j^+ \right ) \;  , \qquad
\xi_i \equiv \frac{i}{\hbar} \sum_{j(\neq i)}^N \left ( 2c_{ij} S_j^z \; -\;
\frac{1}{2} \; a_{ij} S_j^- + 2b_{ij} S_j^+ \right ) \; ,
\ee
where
\be
\label{47}
a_{ij} \equiv D_{ij}^{zz} \; , \qquad 
b_{ij} \equiv \frac{1}{4} \left ( D_{ij}^{xx}- D_{ij}^{yy} - 2i D_{ij}^{xy} \right ) \; , 
\qquad 
c_{ij} \equiv \frac{1}{2} \left ( D_{ij}^{xx} - i D_{ij}^{yz} \right ) \; .
\ee

In the evolution equations, as is well known, there arise pair spin 
correlators that need to be decoupled for obtaining a closed system of
differential equations. For different clusters, we use the semiclassical 
decoupling
\be
\label{48}
 \lgl S_i^\al S_j^\bt \rgl = \lgl S_i^\al \rgl \lgl S_j^\bt \rgl 
\qquad (i \neq j) \;  ,
\ee
supplemented by the account of quantum spin correlations yielding the
appearance of the dephasing term $\gamma_2$. And the spin correlators
for different components of the same cluster, similar to Eq. (13), are 
decoupled as
\be
\label{49}
  \lgl S_j^\al S_j^\bt + S_j^\bt S_j^\al \rgl =
\left ( 2\; - \; \frac{1}{S} \right ) \lgl S_j^\al \rgl \lgl S_j^\bt \rgl 
\qquad ( \al \neq \bt) \;  ,
\ee
in order to retain the correct limiting expressions for spin one-half and
large spins [32-34]. The angle brackets, as earlier, imply statistical 
averaging over the initial statistical operator. This type of spin 
decoupling will lead to the appearance of the expressions
\be
\label{50}
 \xi_0 \equiv \frac{1}{N} \sum_{j=1}^N \; \lgl \xi_j^0 \rgl \; , \qquad
\xi \equiv \frac{1}{N}  \sum_{j=1}^N \; \lgl \xi_j \rgl \; .
\ee
 
As we see, the consideration of an ensemble of magnetic nanoclusters with
dipolar interactions is essentially more complicated than that of a single
nanocluster, treated in the previous sections.

\section{Magnetization Reversal in Ensemble of Nanoclusters}

The quantities of interest are the average spin components, for which it is
convenient to introduce, instead of Eqs. (11) and (12), the relative 
transverse component
\be
\label{51}
u  \equiv \frac{1}{SN} \sum_{j=1}^N \; \lgl S_j^- \rgl
\ee
and the relative longitudinal component
\be
\label{52}
 s  \equiv \frac{1}{SN} \sum_{j=1}^N \; \lgl S_j^z \rgl \; .
\ee
The effective force acting on a cluster spin, instead of Eq. (18), now is
\be
\label{53}
 F = - \; \frac{i}{\hbar} \; \mu_0 H + \om_1 + 
\frac{i}{2} \; \om_2 u^* + \xi \;  .
\ee

Writing down the equations of motion for the spin operators and averaging 
them [46], we come to the evolution equations for the mean spin components
(51) and (52) in the form
\be
\label{54}
 \frac{du}{dt} = - i ( \Om + \xi_0 - i\Gm_2 ) u + Fs \; , \qquad
\frac{ds}{dt} = - \; \frac{1}{2} \left ( u^* F + F^* u \right ) -
\gm_1 ( s - \zeta) \;   ,
\ee
where
\be
\label{55}
 \Gm_2 = \gm_2 \left ( 1 - s^2 \right ) \; , \qquad
\gm_2= \frac{\rho\mu_0^2 S}{\hbar} = \rho\hbar\gm_S^2 S \;  .
\ee
Equations (54) for a nanocluster system replace Eqs. (19) and (20) for a 
single cluster. These equations are to be considered together with Eq. (4)
for the resonator feedback field, with
\be
\label{56}
 m_x = \frac{1}{2} \; \rho \mu_0 S ( u^* + u ) \;  .
\ee
  
Instead of the feedback rate (25) for a single cluster, for $N$ clusters, 
we now have
\be
\label{57}
 \gm_0(N) = \pi N \; \frac{\mu_0^2 S}{\hbar V_{res} } =
\pi \eta \rho\hbar \gm_S^2 S \;  .
\ee
And instead of the coupling function (27), we get
\be
\label{58}
 \al(N) = g \gm_2 ( 1 - As) \left ( 1 - e^{-\gm t} \right ) \;  ,
\ee
with the dimensionless coupling parameter
\be
\label{59}
g \equiv \frac{\gm\om_0\gm_0(N)}{\gm_2(\gm^2+\Dlt^2) } 
\ee
and the effective anisotropy parameter
\be
\label{60}
 A \equiv \frac{\om_A}{\om_0} = \frac{2\om_D + \om_2}{2\om_0} \;  .
\ee
Notice that, for $\pi \eta \sim 1$, the feedback rate (57) is of order
of $\gamma_2$. Under good resonance, with a small detuning $\Delta  \sim 0$,
the coupling parameter (59) is $g \sim \omega_0/ \gamma$.

The system of equations (54) and (21) can be solved numerically, either 
directly, as has been done for magnetic molecules [47], or invoking the 
averaging techniques [39,40], when fast oscillations are averaged out, so 
that the resulting curves are smoothed. Both these methods give close results.
The averaging techniques provide more physically transparent description of 
the initial stage of spin motion, showing that this motion starts with 
stochastic spin fluctuations caused by nonsecular terms of dipolar 
interactions. Thus, dipolar interactions play at the initial stage the positive 
role of a triggering mechanism initiating spin motion, while at the later stage
they play the negative role, by destroying the coherence of spin rotation.

In Fig. 7, we show the results of numerical calculations, involving the 
averaging techniques, for the time dependence of the reduced spin polarization 
(52) for different system parameters. Time is measured in units of 
$\gamma_2^{-1}$ and all frequencies, in units of $\gamma_2$. The realistic value 
for the anisotropy parameter (60) is taken as $A = 0.1$. It is worth 
emphasizing that for $\gamma_2 > \omega_1$, dipolar interactions initiate spin 
dynamics so that $\omega_1$ plays a minor role. In order to stress the 
triggering role of the dipolar interactions, the transverse frequency is set 
to zero, $\omega_1 = 0$. Analogously to Eqs. (38), the initial conditions are
$$
w_0 \equiv w(0) = 0 \;  , \qquad s_0 \equiv s(0) = 1 \; ,
$$
where $w = |u|^2 = x^2 + y^2$. The figure shows that ultrafast magnetization 
reversal happens also in a system of nanoclusters interacting through dipolar 
forces. Even more interesting is the fact that the dipolar interactions play 
the role of a triggering mechanism starting spin dynamics. The magnetization 
reversal is realized during the time of order $1/g \gamma_2 = T_2 / g$, which, 
for $g > 1$ is shorter that the dephasing time $T_2 = 1/\gamma_2$. Hence it is 
feasible to find the system parameters, when dipolar interactions do not disturb 
the coherence of spin motion, provided the sample is coupled to a resonator. 
Without the latter, the Purcell effect does not exist and the ultrafast 
magnetization reversal is impossible.

\section{Discussion}

We have suggested a method for realizing an ultrafast magnetization reversal 
of nanoclusters. The possibility of such a fast reversal is important for a 
number of applications, e.g., for the functioning of various magneto-electronic 
devices, spintronics, magnetic recording and storage, and other information 
processing techniques [48-50]. The idea of the method is based on the coupling 
of the nanocluster with a resonant electric circuit. This is easily achievable 
by placing the nanoclusters inside a magnetic coil. Then the motion of the 
nanocluster magnetic moment produces electric current in the circuit, which 
creates magnetic field acting back on the cluster magnetization. This feedback 
field of the resonator accelerates the magnetization reversal. The reversal 
time can be made many orders shorter than the natural relaxation time.  

First, we have considered a single nanocluster, which makes it possible to 
avoid complications due to distributions of particle sizes, shapes, spin values,
and so on, which could arise in the case of an assembly of many nanoclusters.
In the latter case, the basic complication is the necessity of taking into 
account dipole interactions between the clusters. All these additional 
problems are avoided when dealing with a single cluster.

The case of an ensemble of nanoclusters, interacting through dipolar forces 
is also analyzed. The ultrafast magnetization reversal is feasible for this 
case as well. The reversal occurs during the time shorter than the dipole 
dephasing time, because of which dipolar interactions do not destroy coherent
spin motion that is responsible for the ultrafast reversal. Even more, dipolar
interactions are useful at the initial stage, when they trigger spin dynamics.      

To give the feeling of typical values for the characteristic parameters, let
us make estimates for some nanoclusters. Actually, the family of magnetic 
nanoclusters is very wide and these can display rather different properties
[1-3,51-54]. To be concrete, let us keep in mind the values typical of Co, Fe,
and Ni nanoclusters. The coherence radius for these clusters, below which they 
are in a single-domain state and can display coherent rotation of magnetization 
is $R_{coh} \sim 10$ nm. The standardly formed clusters have the radii 
$R \sim 1-3$ nm. The corresponding cluster volume is $V_1\sim 10^{-20}$ cm$^3$. 
A cluster contains about $N_1 \sim 10^3$ atoms. The atomic density in
a cluster is $\rho_1 \sim 10^{23}$ cm$^{-3}$. The cluster spin is proportional 
to the number of atoms in the cluster, hence $S \sim 10^3$, that is, the magnetic
moment is of the order $10^3 \mu_B$. The magnetic anisotropy parameters (2) are
$K_1 \sim K_2 \sim 10^6$ erg/cm$^3$. The fourth-order anisotropy is much smaller,
$K_4 \sim 10^5$ erg/cm$^3$. This, for the anisotropy parameters of Hamiltonian (1), 
gives $D \sim D_2 \sim 10^{-20}$ erg. And for the fourth-order anisotropy, this 
would make $D_4 \sim 10^{-27}$ erg.

These values, for the anisotropy frequencies (14) to (16) yield 
$\omega_D \sim \omega_2 \sim \omega_A \sim 10^{10}$ Hz. This corresponds to the 
anisotropy field $B_A \equiv \omega_A / \gamma_S \sim 10^3$ G. The Zeeman frequency,
for the magnetic field $B_0 \sim 1$ T is $\omega_0 \sim 10^{11}$ Hz. Note that
the present day facilities allow for the generation of magnetic fields as high as
about $100$ T [55]. The feedback rate (25) is $\gamma_0 \sim 10^{10}$ s$^{-1}$.
This rate provides the reversal time $t_{rev} \sim 1/ \gamma_0 \sim 10^{-10}$ s. 
  
The blocking temperature, below which thermally activated reversals are exponentially 
suppressed is $T_B \sim 10-40$ K. The typical prefactor in the Arrhenius law is
$\gamma_A \sim 10^{9} - 10^{11}$ s$^{-1}$. The anisotropy energy barrier in the 
Arrhenius law is $E_A \sim 10^{-14}$ erg. This gives the anisotropy temperature 
$E_A/k_B \sim 100$ K. The resulting relaxation time $T_1 \equiv 1/ \gamma_1$, below 
the blocking temperature, is rather long. Thus, even at the temperature $T = 10$ K, 
we have $T_1 \sim 10^{-5}$ s. At the temperature $T = 5$ K, one has $T_1 \sim 0.1$ s.
And for temperature $T = 1$ K, the thermal reversal time is astronomically large,
being $T_1 \sim 10^{34}$ s. But, coupling the nanocluster to a resonator, produces
a very short reversal time $t_{rev} \sim 10^{-10}$ s, independently of the value $T_1$.
The reversal time could be made shorter by choosing the appropriate types of 
nanoclusters and resonator properties.

As has been explained above, the spin relaxation in the system of nanoclusters, 
which would be caused by the photon exchange between different spins is 
negligible [33,34,43,44]. The corresponding radiation width is
$$
 \gm_{rad} = \frac{2\om^3\mu_0^2 S}{3\hbar c^3} = 
\frac{2\om^3}{3\rho c^3}\; \gm_2 \;   .
$$
For the typical density of nanoclusters $\rho \sim 10^{20}$ cm$^{-3}$, the spin
of a cluster $S \sim 10^3$, and frequency $\omega \sim 10^{11}$ Hz, we get 
$\gamma_{rad} \sim 10^{-8}$ s$^{-1}$, which is much smaller than the dipolar
width $\gamma_2 \sim 10^{10}$ s$^{-1}$. Therefore the relaxation time, due to
the photon exchange between spins, 
$t_{rad} = 1/ \gamma_{rad} \sim 10^8$ s $\sim 10$ years, is
enormously larger than the dipolar dephasing time $T_2 \sim 10^{-10}$ s. This
confirms that the photon exchange mechanism plays no role in the spin relaxation,
that is, the Dicke effect for spin systems does not exist. But the ultrafast 
magnetization reversal is completely due to the Purcell effect. 

The typical density of an ensemble of clusters is $\rho \sim 10^{20}$ cm$^{-3}$.
With the natural dipolar width $\gamma_2 \sim 10^{10}$ s$^{-1}$, the dephasing 
time is $T_2 \sim 10^{-10}$ s. As is seen in Fig. 7, the reversal time
can be an order shorter than the dephasing time, being $t_{rev} \sim 10^{-11}$ s.

Concluding, by coupling nanoclusters to a resonant electric circuit, it is 
possible to realize ultrafast magnetization reversal for single nanoclusters
as well as for assemblies of nanoclusters. For such nanoclusters as Fe, Co, or Ni,
the reversal time can be made as short as $10^{-11}$ s.

\vskip 2mm

{\bf Acknowledgement}

The authors acknowledge financial support from the Russian Foundation for Basic 
Research.

\newpage

\begin{center}
{\Large{\bf Figure Captions}}
\end{center}

\vskip 3cm

{\bf Fig. 1}. Solutions to the evolution equations: (a) $x=x(t)$; (b) $y=y(t)$;
(c) $s=s(t)$; (d) $h=h(t)$ for the parameters $\omega_0 = \omega = 10$,
$\omega_D = \omega_1 = \omega_2 = 1$, $\gamma = 1, \gamma_1 = 10^{-3}$.
The parameters are measured in units of $\gamma_0$ and time, in units
of $1/\gamma_0$. To emphasize the role of the resonator feedback field,
the solutions in the presence of the resonator (solid lines) are compared
with those for the case of no resonator (dashed lines).

\vskip 1cm

{\bf Fig. 2}. Role of the resonator attenuation. Magnetization as a function of 
time for the parameters $\omega_0=\omega=10$, $\omega_D=\omega_1=\omega_2=1$, 
and varying resonator attenuation: (a) $\gamma=0.1$; (b) $\gamma=1$ (solid line) 
and $\gamma = 10$ (dashed line).

\vskip 1cm

{\bf Fig. 3}. Role of the Zeeman frequency. Magnetization as a function of time 
for the parameters $\omega_D = \omega_1 = \omega_2 = 1$, $\gamma = 1$, and 
varying Zeeman frequency: (a) $\omega_0 = \omega = 1$ (solid line) and 
$\omega_0 = \omega = 10$ (dashed line); (b) $\omega_0 = \omega = 100$.

\vskip 1cm

{\bf Fig. 4}. Role of the triggering field. Magnetization as a function of time 
for the parameters $\omega_0=\omega=10$, $\omega_D = \omega_2 = 1$, and varying 
triggering field: (a) $\omega_1 = 0.001$ (solid line) and $\omega_1 = 1$ 
(dashed line); (b) $\omega_1 = 10$.

\vskip 1cm

{\bf Fig. 5}. Role of the anisotropy. Magnetization as a function of time  for 
the parameters $\omega_0 = \omega = 10$, $\omega_1 = 1, \gamma = 1$, and varying
anisotropy frequencies $\omega_D = \omega_2 = 1$ (dashed-doted line);
$\omega_D = \omega_2 = 10$ (solid line); $\omega_D = \omega_2 = 15$ (dashed 
line).

\vskip 1cm

{\bf Fig. 6}. Role of the resonance. Magnetization as a function of time for 
the parameters $\omega_0=10$, $\omega_D=\omega_1=\omega_2=1$, $\gamma = 1$,
and varying detuning from the resonance, with $\omega = 1$ (solid line);
$\omega = 10$ (dashed-dotted line); $\omega = 20$ (dashed line).

\vskip 1cm

{\bf Fig. 7}. Magnetization reversal in the system of nanoclusters with
dipolar interactions. The system parameters are $\gm_1=10^{-3}$, $\Dlt=0$,
$A = 0.1$, $\om_1 = 0$. Time is measured in units of $T_2\equiv1/\gm_2=10^{-10}$ 
s, and the frequencies, in units of $\gm_2$. Other parameters are: 
$\gm=10$, $\om_0=\om=1000$, $g=100$ (solid line); $\gm=1$, $\om_0=\om=100$, 
$g=100$ (dashed-dotted line); $\gm=10$, $\om_0=\om=100$, $g=10$ (dashed line). 
The shown functions of time are: (a) coherence intensity $w(t)$; (b) reduced
magnetization $s(t)$.

\newpage

\begin{figure}[ht]
\vspace{9pt}
\centerline{
\hbox{ \includegraphics[width=16cm]{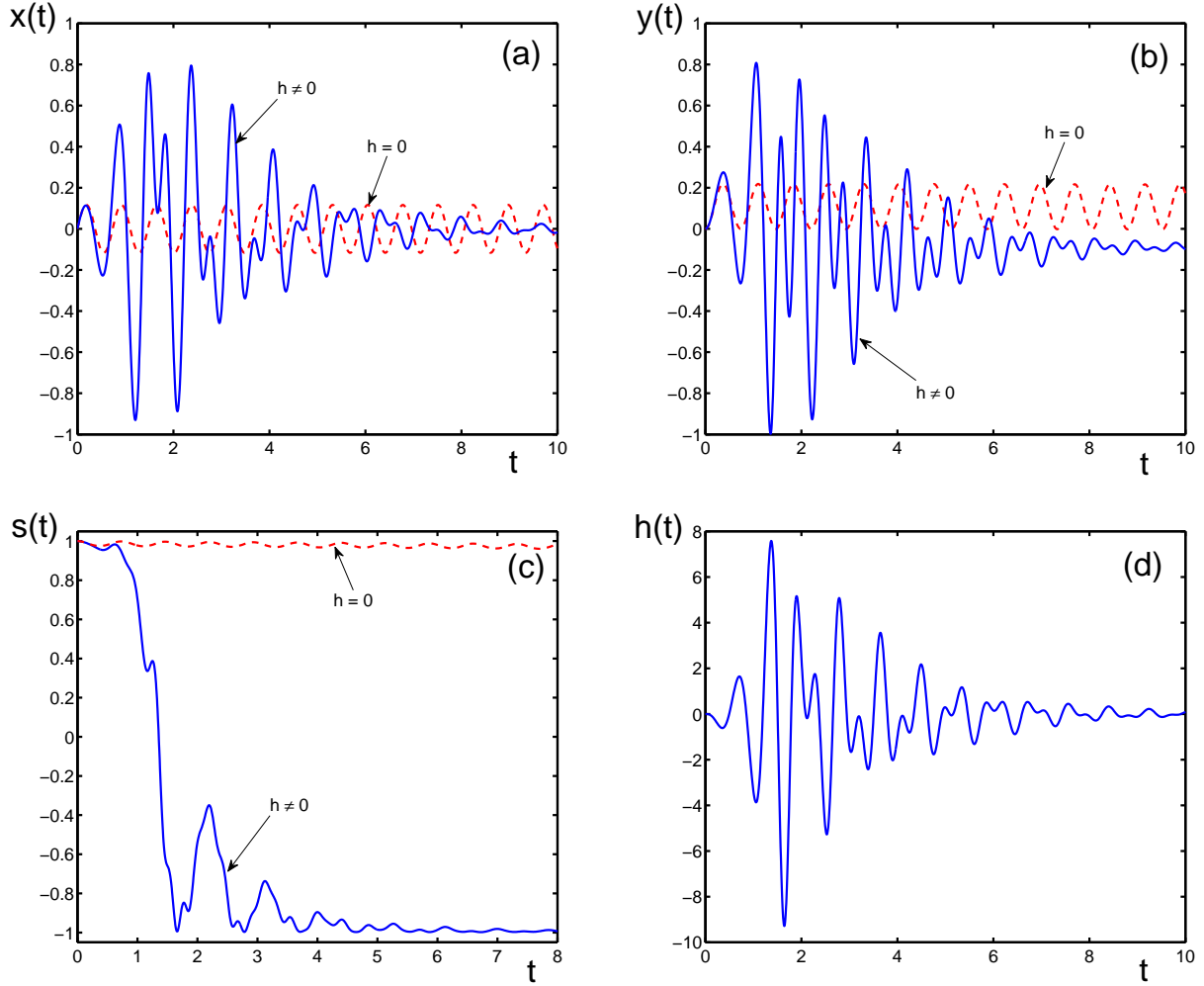} } }
\caption{Solutions to the evolution equations: (a) $x=x(t)$; (b) $y=y(t)$;
(c) $s=s(t)$; (d) $h=h(t)$ for the parameters $\omega_0 = \omega = 10$,
$\omega_D = \omega_1 = \omega_2 = 1, \gamma = 1, \gamma_1 = 10^{-3}$.
The parameters are measured in units of $\gamma_0$ and time, in units
of $1/\gamma_0$. To emphasize the role of the resonator feedback field,
the solutions in the presence of the resonator (solid lines) are compared
with those for the case of no resonator (dashed lines).}
\label{fig:Fig.1}
\end{figure}

\begin{figure}[ht]
\vspace{9pt}
\centerline{
\hbox{ \includegraphics[width=16cm]{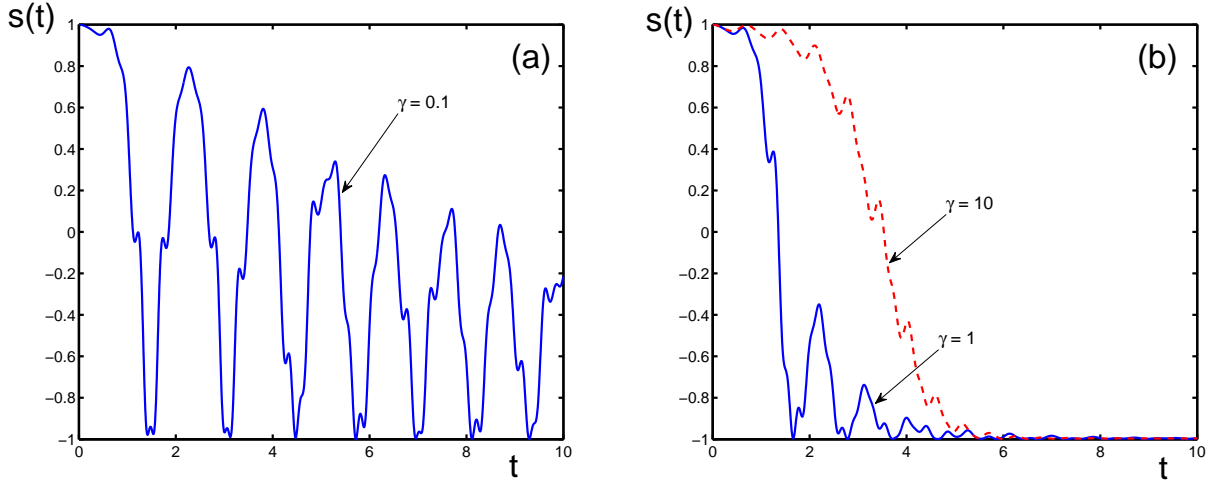}  } }
\caption{Role of the resonator attenuation. Magnetization as a function of 
time for the parameters $\omega_0 = \omega = 10, \omega_D = \omega_1 = \omega_2 = 1$ 
and varying resonator attenuation: (a) $\gamma = 0.1$; (b) $\gamma = 1$ (solid line) 
and $\gamma = 10$ (dashed line).}
\label{fig:Fig.2}
\end{figure}

\begin{figure}[ht]
\vspace{9pt}
\centerline{
\hbox{ \includegraphics[width=16cm]{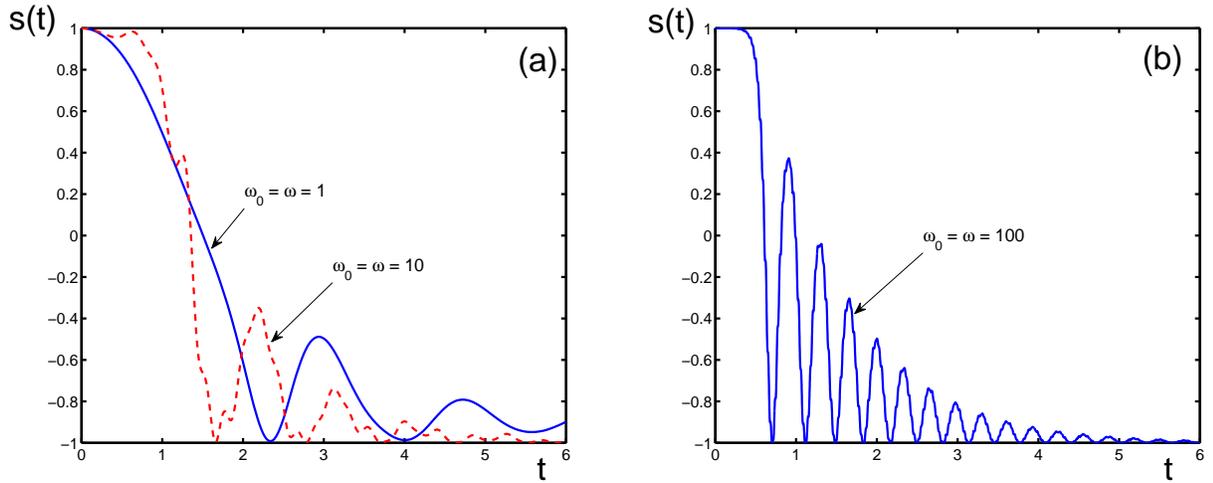} } }
\caption{Role of the Zeeman frequency. Magnetization as a function of time 
for the parameters $\omega_D = \omega_1 = \omega_2 = 1$, $\gamma = 1$ and 
varying Zeeman frequency: (a) $\omega_0 = \omega = 1$ (solid line) and 
$\omega_0 = \omega = 10$ (dashed line); (b) $\omega_0 = \omega = 100$.}
\label{fig:Fig.3}
\end{figure}

\begin{figure}[ht]
\vspace{9pt}
\centerline{
\hbox{ \includegraphics[width=16cm]{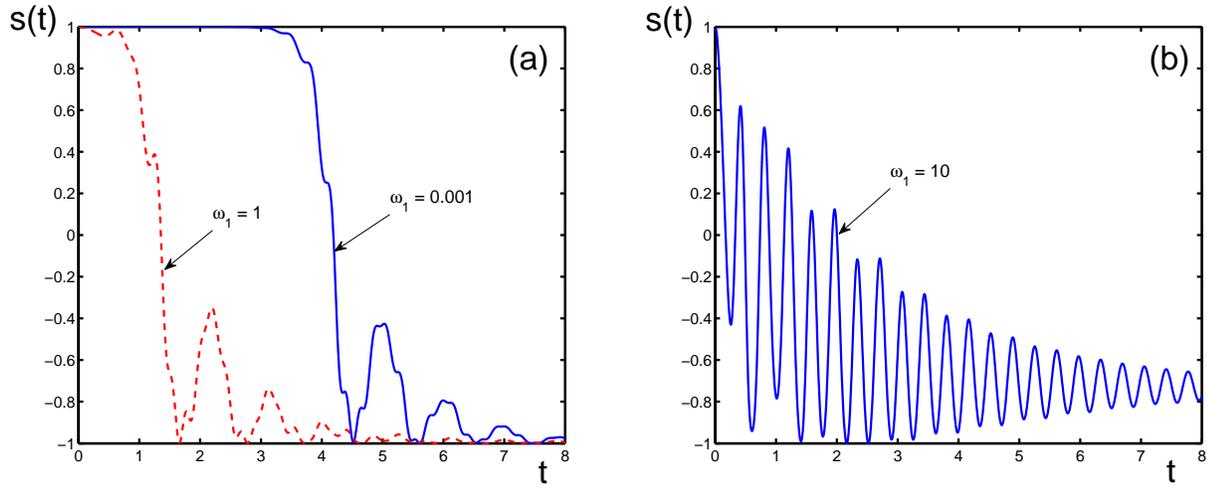} } }
\caption{Role of the triggering field. Magnetization as a function of time for 
the parameters $\omega_0=\omega=10$, $\omega_D=\omega_2=1$, and varying 
triggering field: (a) $\omega_1 = 0.001$ (solid line) and $\omega_1 = 1$ 
(dashed line); (b) $\omega_1 = 10$.}
\label{fig:Fig.4}
\end{figure}

\begin{figure}[ht]
\vspace{9pt}
\centerline{
\hbox{ \includegraphics[width=8cm]{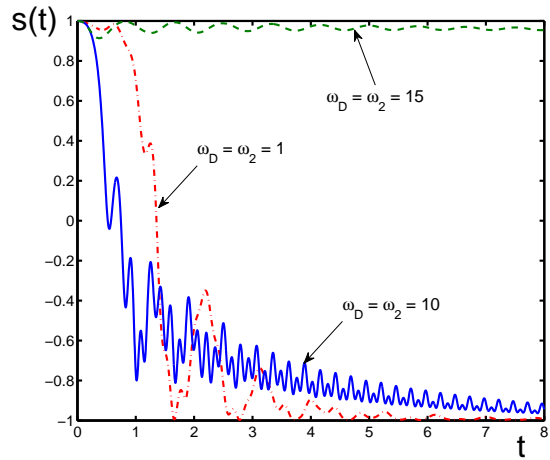}  } }
\caption{Role of the anisotropy. Magnetization as a function of time  
for the parameters $\omega_0=\omega=10$, $\omega_1=1$, $\gamma = 1$, and 
varying anisotropy frequencies $\omega_D = \omega_2 = 1$ (dashed-doted 
line); $\omega_D=\omega_2=10$ (solid line); $\omega_D = \omega_2 = 15$ 
(dashed line).}
\label{fig:Fig.5}
\end{figure}

\begin{figure}[ht]
\vspace{9pt}
\centerline{
\hbox{ \includegraphics[width=8cm]{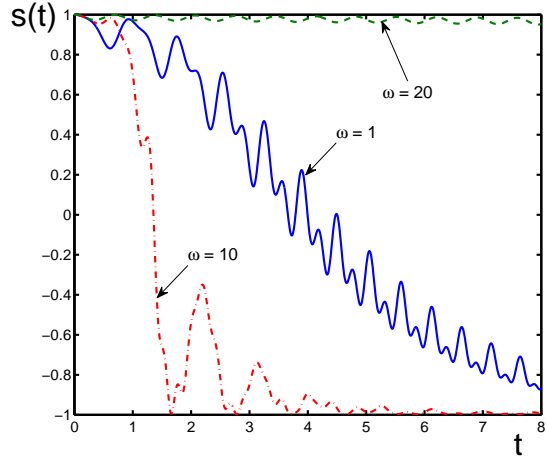}  } }
\caption{ Role of the resonance. Magnetization as a function of time for 
the parameters $\omega_0 = 10$, $\omega_D =\omega_1 =\omega_2 = 1$, 
$\gamma = 1$, and varying detuning from the resonance, with $\omega = 1$ 
(solid line); $\omega=10$ (dashed-dotted line); $\omega=20$ (dashed line).}
\label{fig:Fig.6}
\end{figure}

\vskip 3cm

\begin{figure}[ht]
\vspace{9pt}
\centerline{
\hbox{ \includegraphics[width=16cm]{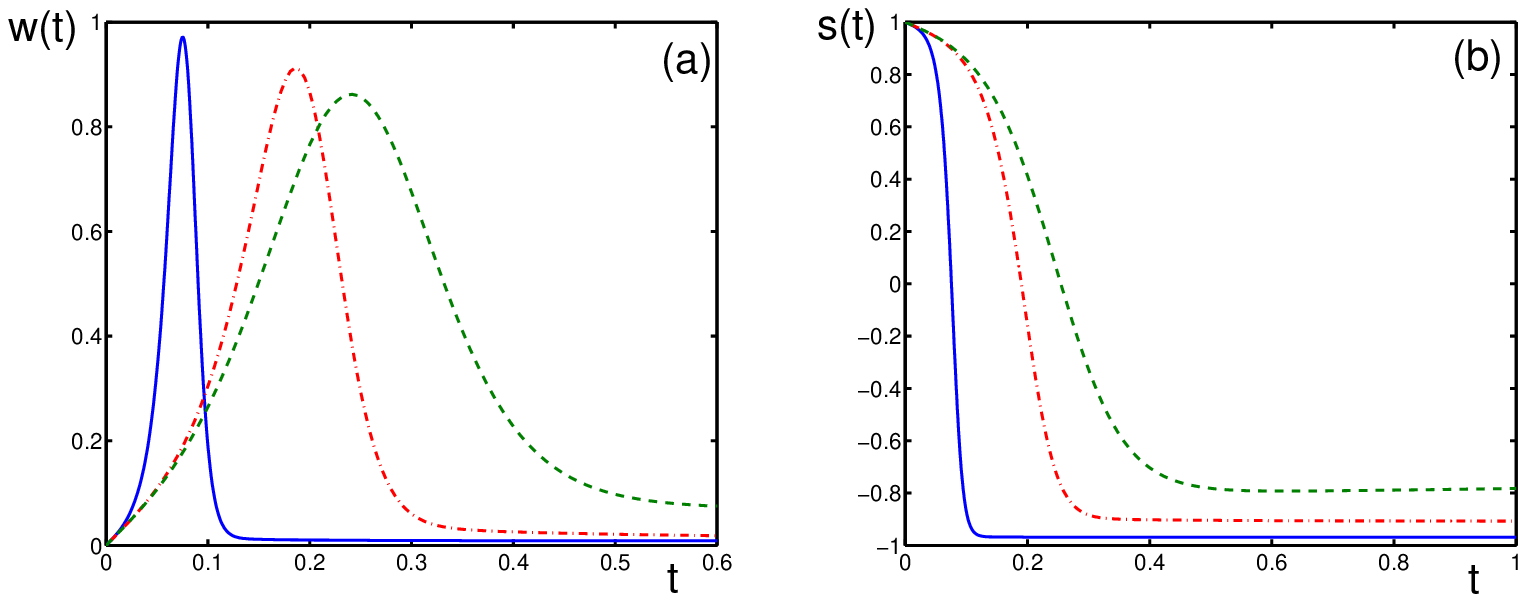}  } }
\caption{Magnetization reversal in the system of nanoclusters with
dipolar interactions. The system parameters are $\gm_1=10^{-3}$, $\Dlt=0$,
$A = 0.1$, $\om_1 = 0$. Time is measured in units of $T_2\equiv1/\gm_2=10^{-10}$ 
s, and the frequencies, in units of $\gm_2$. Other parameters are: 
$\gm=10$, $\om_0=\om=1000$, $g=100$ (solid line); $\gm=1$, $\om_0=\om=100$, 
$g=100$ (dashed-dotted line); $\gm=10$, $\om_0=\om=100$, $g=10$ (dashed line). 
The shown functions of time are: (a) coherence intensity $w(t)$; (b) reduced
magnetization $s(t)$.}
\label{fig:Fig.7}
\end{figure}

\end{document}